\documentstyle[aps,preprint,epsf]{revtex}

\begin{document}   

\title{A transient network of telechelic polymers and microspheres : structure and rheology}

\author{F. Molino$^{1}$}
\author{J. Appell$^{1}$, M. Filali$^{2}$, E. Michel, G. Porte$^{1}$ }
\author{S. Mora, E. Sunyer$^{1}$}

\address{$^{1}$ Groupe de Dynamique des Phases Condens\'ees, Unit\'e Mixte de Recherche CNRS/Universit\'e Montpellier II n$^0$ 5581, F-34095 Montpellier, Cedex 05, France\\
$^{2}$ Laboratoire de Physique du Solide,Facult\'e des sciences Dharmehraz, BP1796 Atlas,FES, Morocco}

\maketitle

\begin{abstract} 
We study the structure and dynamics of a transient network composed of droplets of microemulsion connected by telechelic polymers. The polymer induces a bridging attraction between droplets without changing their shape. A viscoelastic behaviour is induced in the initially liquid solution, characterised in the linear regime by a stretched exponential stress relaxation. We analyse this relaxation in the light of classical theories of transient networks. The role of the elastic reorganisations in the deformed network is emphasized. In the non linear regime, a fast relaxation dynamics is followed by a second one  having the same rate as in the linear regime. This behaviour, under step strain experiments, should induce a non monotonic behaviour in the elastic component of the stress under constant shear rate. However, we obtain in this case a singularity in the flow curve very different from the one observed in other systems, that we interpret in terms of fracture behaviour.
\end{abstract}

\newpage

\section{introduction}
The study of {\em transient networks} has been renewed by investigations of systems of {\em telechelic polymers}, modified at both extremities with end groups differing in affinity from the main chain (\cite{1}). To obtain a so-called {\em transient network}, one starts from a monodisperse oil/water droplets microemulsion of oil droplets in water, stabilised at a fixed radius by surfactant coating of the droplets, and add  telechelic polymers whose extremities will preferentially fix to the droplets (\cite{2,3,4}). In this paper  we report first on the influence of the added polymer to the structure of the equilibrium solution at various concentrations. Then we study the dynamics of such a system under shear. 
\section{System description : structure and interaction}
\subsection{Experimental}
We first obtain a thermodynamically stable oil/water microemulsion droplets, and then add the hydrophobically endcapped polymers. Cetyl pyridinium chloride (CPCL) and octanol were used as surfactant and cosurfactant to stabilise an emulsion of decane in brine. The spontaneous radius of the surfactant film is adjusted by the surfactant to cosurfactant ratio. The decane is added up to a value just below the line of emulsification failure. For a typical volume concentration in oil drpolets $\phi=$10\%, the mass ratios were 4.5\% of decane, 1.6\% of octanol, and 6.3\% of CPCL, corresponding to a cosurfactant/surfactant ratio of 0.25. Under these conditions, it is well known (\cite{3bis}) that one obtains a {\em  stable dispersion of monodisperse droplets}.\\
The associative polymers consist of a poly(ethylene-oxide) (PEO) chain chemically modified at both extremities by addition of a C$_{18}$ alkyl group.\\
\subsection{Phase behaviour and structure}
 The telechelic polymers introduces an effective attraction between the micelles through bridging. The relevant parameter is in this respect the number $r$ of alkyl goups  per droplet. Below $\Phi=$10\%, the initially clear and homogeneous microemulsion undergo a phase separation upon addition of polymers.\\
Small angle neutron scattering data have been obtained at LLB-Saclay, on the PACE spectrometer. Neutrons are mainly scattered by the hydrophobic cores. From these studies (see \cite{4}), the average radius is $R=62 \mbox{\AA}$.  It is also apparent that the associative polymer has no influence on the droplets shape.\\
At small concentrations, the addition of polymer seems to induce both the apparition of a correlation peak in neutrons 9see \cite{4}), and a strong diffusion at small angles. In this case the droplets are induced to spend a longer time at a {\em preferential distance from each other, fixed by the polymer end-to-end distance}, during a typical {\em bridge life time}, due to polymer attraction (see \cite{4}). This effective attraction accounts for the increase in osmotic compressibility and hence to the small $q$ enhanced scattering, as well as for the phase separation. At high concentrations, the bridging has no other effect than strenghtening the weak interaction between the droplets. Slightly below  10\%, there exist a balance concentration for which the polymer natural lenght exactely matches the constraint of liquid-like homogeneity for the droplets.\\
\section{Dynamical properties and rheology}
\subsection{Introduction}
We select a concentration of 10\%, in which the droplets remain statistically most of the time at distances slightly below the average end to end distance of the polymer. Thus we can study the influence of $r$ on a  large range (1 to 20) without phase separation. Rheological measurments have been performed both in stress controlled and strain controlled setups. The usual setup was cone and plate \footnote{The rheometers used respectively in stress-controlled and strain-controlled experiments were a Physica US200, and a Rheometrics RFS II} . 
\subsection{Linear behaviour and stress relaxation}
 Step-strain experiments were performed on the system. In figure 1, we show typical responses for strains varying from 50\% to 400\%. For small deformations the response for all $r$ values takes the form of a slightly stretched exponential, $G(t)=G_{0r}e^{-(\frac{t}{\tau_r})^{0.8\pm 0.05}}$ with a value of the exponent independent of $r$. The elastic modulus $G_{0r}$ is  in contrast strongly $r$-dependent, as can be seen in figure 2.\\
The elastic modulus goes to {\em zero} for a finite value of $r$, together with the time $\tau_r$. This is clearly a {\em percolation effect}. The elastic stress being transported through the sample by the polymers acting as hookean springs, an infinite geometrical path of strings should exist (at least) in the system for a non-zero modulus to be observed. Indeed the $r$ dependence of $G_{0r}$ follows very closely the power-law behaviour $G_{0r}\sim(r-r_p)^{1.7}$ predicted years ago from an  analogy with electrical networks (see \cite{7}). The $r=6$ value is confirmed by independent observations of the relaxation time by dynamic light scattering. The fraction of polymer forming {\em loops} instead of {\em bridges} is  unknown. If the number of bridges per droplet at the percolation point was 1.5, as expected from the geometrical theory of percolation (see \cite{8}), the rate of bridges to loops would be around 25\%.\\ 
The typical scale of the relaxation time should be fixed by the {\em average life time of a bond}  related to the energy necessary to break the bond  by an Arrhenius-like law  $\tau=\frac{1}{\nu}e^{(\frac{W}{k_B T})}.$ (see \cite{9}). The stress is relaxed as the {\em loaded} bridges break and reform. {\em But they reform in an unloaded state}, and henceforth do not participate anymore to the stress if the sample is no more macroscopically deformed after $t=0$. In this picture, at any time $t$, the number of chains still loaded can be described by an effective connectivity $r(t)$, whose evolution is  $r(t) \simeq r_0 e^{-(\frac{t}{\tau})}$.  From the modulus dependence {\em at fixed r} an evolution law follows (\cite{10}) : $G(t)\sim(r(t)-r_p)^{1.7}\sim(r_0 e^{-(\frac{t}{\tau})}-r_p)^{1.7}.$ This is plainly wrong as can be seen from figure 3a : the predicted behaviour is not a streched exponential : {\em the modulus should go to zero in a finite time} as $r(t)$ reaches $r_p$! \\
Two assumptions underlies this prediction : first, that links are {\em randomly broken} in the percolating network. This would be wrong if the network was {\em not deformed in an affine way}, some bridges being more streched locally. The second hypothesis is the elimination in the stress calculations of all the broken bridges, although they are reconnected. This assumption, from the original {\em Green-Tobolski} model (\cite{11,12}), has been central in all transient network theories. It  could be challenged by the observation that the reconnected bridges, even if they cannot participate immediately to the stress, due to the isotropic character of their distribution, could later participate again in the stress relaxation, due to {\em rearrangements of the droplets}. Upon breaking of a bond, the droplets will relax to a position governed on average by their simple mechanical equilibrium under the springs forces. This will occur on a time short compared to the duration of a bond. During these droplets relaxations, polymers reconnected earlier in a statistically isotropic unloaded state will again carry some load. Thus the finite time percolation breaking is prevented by the loading of new bridges.  We  show in figure 3b results from 2D off-lattice numerical simulations in which the network has been allowed or not to relax. Moreover the mechanisms of breaking and reconnections have been introduced following a realistic prescription. The simulations support the view that the contribution of the reconnected bridges to the stress can prevent the percolation failure at a finite time and lead to an exponential relaxation.\\ 
\subsection{Non linear regime}
When the amplitude of the strain is increased, the stress relaxation transforms at a critical $\gamma_f$ value independent of $r$ into a two steps process : an initial fast exponential decay ($\sim$ 10\% of the linear time) is followed by a linear behaviour. The life time of a bridge depend on the stretching of the polymer chain, whose stored elastic energy can supply through mechanical work a part of the activation energy  : the time $\tau$ becomes $\tau=\frac{1}{\nu}e^{(\frac{W-fa}{k_B T})},$ where $f$ is the elastic {\em force} exerted by the polymer, and $a$ a typical lenght on which the force works.  Thus from an isotropic unloaded distribution, the deformation makes an ellipsoid, which then relaxes (as the stress), with stretched bridges breaking more quickly. In a short time, the system is back in a configuration of unstretched bonds, whose dynamics obeys henceforth the linear rate of breaking. The stress thus always relaxes at long times with a typical rate identical with the one observed in the {\em linear regime}, as can be seen comparing the asymptotic slopes of the curves in figure 1.\\
 \subsection{Flow behaviour}
 The non linear response gives us a hint of the {\em flow behaviour}. Under constant shear rate, only the slowest relaxation processes are relevant. The effective elasticity associated with these, as obtained from the $t=0$ extrapolation of the long time relaxation, decreases strongly as the strain amplitude is increased. Thus a {\em non monotonous stress/rate relation } is expected for the elastic part of the stress since $\sigma (\dot{\gamma}) \sim \sigma_0 (\dot{\gamma}\tau)=\dot{\gamma}\tau G_0(\dot{\gamma}\tau)$ (see \cite{13}). On other systems such as giant micelles, this leads to singularities in the flow curve such as plateaux behaviours (\cite{14}). \\
The picture emerging from the stationary flow curve represented in figure 4 (for a value of $r=12$) is quite different. An initial Newtonian behaviour is abruptely followed by a sudden drop of the stress at a critical shear rate $\dot{\gamma}_c$, with $\dot{\gamma _c} \tau \sim 0.3 $, $\tau$ being the linear time of the same system. A second regime of flow follows, which can be described by a Bingham fluid law $\sigma=\sigma_1 +\eta \dot{\gamma}$. Between these two regimes a $d \sigma /d \dot{\gamma} < 0$ {\em stationary region } is observed. The results compare well to the stress-controlled rheology. Both the relaxation after high deformation and the flow curve suggest that we are observing a {\em fracture behaviour}. The singular drop in stress cannot be acounted for by an underlying non monotonic constitutive equation.  This should lead to a more progressive (continuous) behaviour of $\sigma$. Thus we propose to consider this behaviour in terms of {\em fracture propagation} through the material. We expect that localised fractures appear all the time du to local percolation breakings. The concentration of stress at the boundaries of the fracture can induce the movment of the fracture tip, at a stress-dependent speed. Once a fracture has propagated throughout the geometry of the sample, the stress drops. {\em But when the stress exceeds again the critical value, a second fracture should occur. This does not happen}. We tentatively conclude that  it is a {\em boundary fracture}.  The intermediate stationary region of decreasing stress indicates also that under monitoring of the shear rate in the  $\dot{\gamma}_c$ region, the fracture can heal, the linear (newtonian) behaviour being recovered under decreasing $\dot{\gamma}$.\\
\subsection{Conclusion}
On a model system of transient network, we have demonstrated first the role of {\em percolation} in determining the instantaneous elastic behavior. We then stressed the importance of the {\em elastic reorganisations } of the network, and of the {\em re-loading of the broken bonds}. When the system is highly deformed ($\sim$ 300\%), a {\em fracture behavior} is observed. Many important informations concerning the {\em evolution of the structure} of the material are lacking. We just do not know if the organisation of the droplets or of the polymers changes under flow, or under large deformations. Future X-ray scattering observations or NMR will be the only way to gather further insight into the non linear and flow behaviour.

\begin{figure}
\caption{Strain relaxation $G(t)$ for a $\phi$=10\%, $r$=40 system. The deformations are, from top to bottom curve, 50\%, 100\%, and 400\% respectively. The 50\% curve is in the linear regime and can be fitted by a stretched exponential as indicated in the text.}
\label{fig1}
\end{figure}

\begin{figure}
\caption{Evidence of percolation behaviour : the $r$-dependence of the instantaneous elastic modulus.  The relaxation time goes also to zero for the same value of $r$. Power law behaviours  $G_0\sim(r-r_p)^{1.7}$ and $\tau_r \sim (r-r_p)^{0.8}$ were measured from rheology. From DLS the relaxation time behaviour is $\tau_r \sim (r-r_p)^{1.2}$}
\label{fig2}
\end{figure}

\begin{figure}
\caption{ (3.a) Predicted behaviour of the stress relaxation under the assumptions of the classical transient network theories, compared with actual relaxation. The predicted behaviour goes to zero in a finite time, whereas the actual relaxation decays exponentially. (3.b) Numerical experiments performed in two dimensions on arrays of $\sim$ 1000 droplets, first without taking into account the reconnected bonds in the stress calculation -bottom curve-, then taking these bonds into account and letting the network relax to re-load these bonds -top curve-. }
\label{fig6}
\end{figure}

\begin{figure}
\caption{Stationary flow curve (stress versus shear rate) in both stress and strain controlled rheology. Volume concentrations in oil is $\phi$=10\%, and connectivity is $r$=10.}
\label{fig7}
\end{figure}


\newpage

\begin{thebibliography}{12345 123}
\bibitem{1} M. Winnik, A. Yetka, {\em Curr. Opi. Coll. \& Inter. Sc.}, {\bf 1997}, {\em 2}, 424 
\bibitem{2} Bagger-J\"orgensen {\em et al.}, to appear
\bibitem{3} M. Odenwald {\em et al.}, Macromolecules, {bf 1995}, {\em 28}, 5069 
\bibitem{3bis} S. Safran, Phys. Rev. A, {\bf 1991}, {\em 43}, 2903
\bibitem{4} M. Filali {\em et al.}, J. Phys. Chem. B, {bf 1999}, {\em to appear}
\bibitem{5} B. Cabane, R. Duplessix, Adv. Coll. \& Inter. Sci., {\bf 1982}, {\em 41}, 149 
\bibitem{6} T. Annable {\em et al.}, J. Rheol., {\bf 1993}, {\em 37}, 695
\bibitem{7} P. G. De Gennes, Journal de Physique - Lettres, {\bf 1976}, {\em 37}, L-1
\bibitem{8} D. Stauffer, {\em Introduction to percolation theory}, Taylor \& Francis, {bf 1985}
\bibitem{12} F. Tanaka, S. F. Edwards, {\em Macromolecules}, {\bf 25}, 5495 (1992)
\bibitem{9} M. Green, A. Tobolski, Journ. of Chem. Phys., {bf 1946}, {\em 14}, 80
\bibitem{10} Thanks to {\em Tony Maggs} for this suggestion
\bibitem{11} M. Yamamoto, J. Phys. Soc. Jpn, {\bf 1956}, {\em 11}, 413
\bibitem{13} G. Porte, J. F. Berret, J. Harden, J. Phys II France, {\bf 1997}. {\em 7}, 459
\bibitem{14} J. F. Berret {\em et al.}, Phys. Rev. E, {\bf 1997}. {\em 55}, 1668
\end{thebibliography}
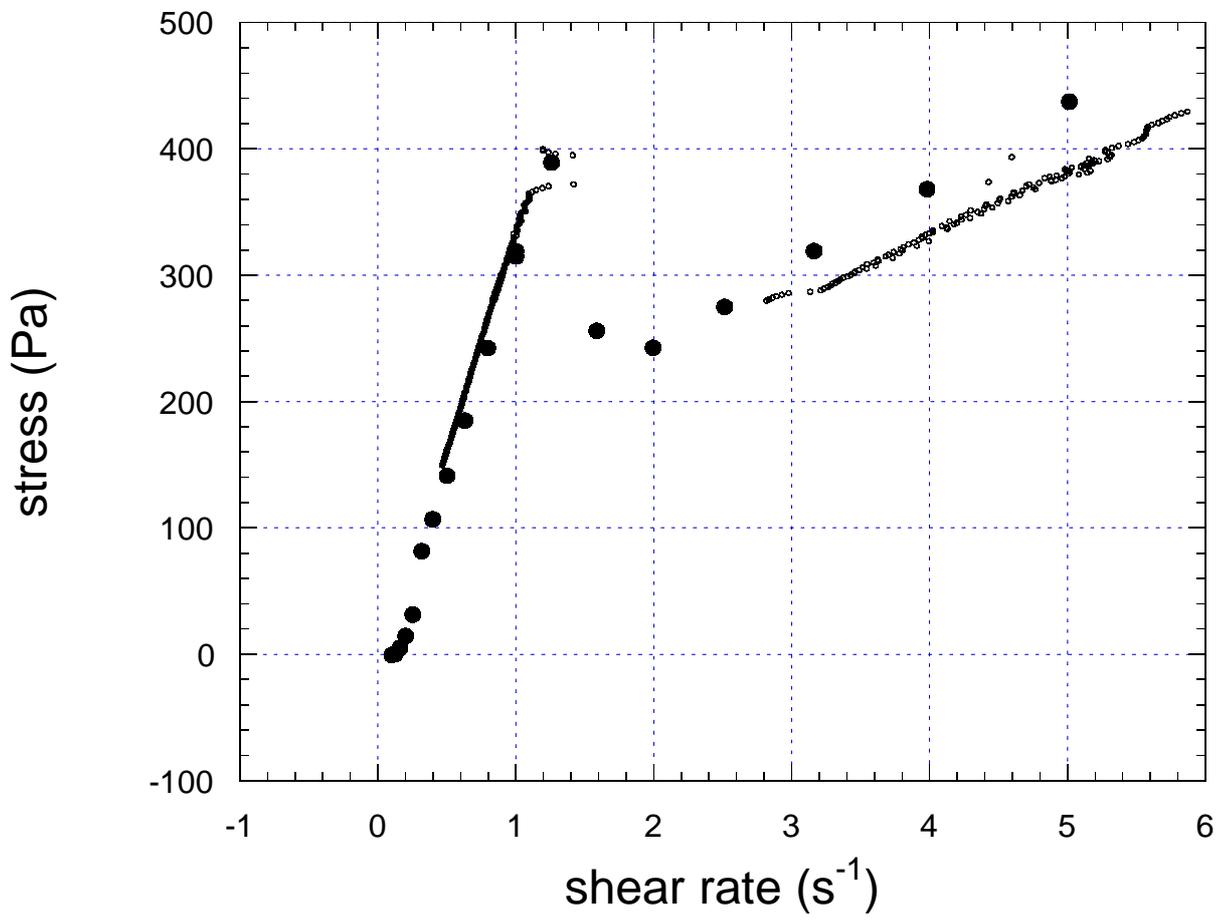
\end{document}